\documentclass[12pt,preprint]{aastex}

\def \kms{km~s$^{-1}$}
\def \coi{$^{12}$CO}
\def \coii{$^{13}$CO}
\def \hi{H{\small I}}
\def \hii{H{\small II}}
\def \nhiii{NH$_3$}

\shorttitle{Arecibo 6.7 GHz Methanol Maser Survey -- III}

\begin{document}
\shortauthors{Pandian, Menten, \& Goldsmith}

\title{The Arecibo Methanol Maser Galactic Plane Survey - III: Distances and Luminosities}
\author{J. D. Pandian\altaffilmark{1}, K. M. Menten\altaffilmark{1} and P. F. Goldsmith\altaffilmark{2}}
\altaffiltext{1}{Max-Planck-Institut f\"{u}r Radioastronomie, Auf dem H\"{u}gel 69, 53121 Bonn, Germany; [jpandian;kmenten]@mpifr-bonn.mpg.de}
\altaffiltext{2}{Jet Propulsion Laboratory, California Institute of Technology, Pasadena, CA 91109; Paul.F.Goldsmith@jpl.nasa.gov}

\begin{abstract}
We derive kinematic distances to the 86 6.7 GHz methanol masers discovered in the Arecibo Methanol Maser Galactic Plane Survey. The systemic velocities of the sources were derived from \coii~($J=2-1$), CS ($J=5-4$), and \nhiii~observations made with the ARO Submillimeter Telescope, the APEX telescope, and the Effelsberg 100 m telescope, respectively. Kinematic distance ambiguities were resolved using \hi~self-absorption with \hi~data from the VLA Galactic Plane Survey. We observe roughly three times as many sources at the far distance compared to the near distance. The vertical distribution of the sources has a scale height of $\sim 30$ pc, and is much lower than that of the Galactic thin disk. We use the distances derived in this work to determine the luminosity function of 6.7 GHz maser emission. The luminosity function has a peak at approximately $10^{-6}~L_\sun$. Assuming that this luminosity function applies, the methanol maser population in the Large Magellanic Cloud and M33 is at least 4 and 14 times smaller, respectively, than in our Galaxy.
\end{abstract}


\section{Introduction}
Methanol masers at 6.7 GHz are unique compared to their OH and H$_2$O counterparts in that they appear to be exclusively associated with early phases of massive star formation. They are hence extremely useful tools to identify and study very young massive star forming regions. To date, over 800 sources have been detected through various targeted and blind surveys of the Galactic plane (e.g. \citealt{gree09}; compilation of \citealt{xu09b}).

In spite of the extensive surveys to date for 6.7 GHz methanol masers, their luminosity function remains largely unknown. Since 6.7 GHz methanol masers are closely associated with mainline OH masers in massive star forming regions, \citet{casw95} suggested that the luminosity function of methanol masers would be similar to that of OH masers, but with a scaling factor to account for the fact that methanol masers are on average much more intense than OH masers. \citet{van96} use a probabilistic approach to assign distances and estimated the luminosity function to have a power-law behavior with an index around --2. Both these studies used only the peak flux density rather than the integrated flux for determining the luminosity functions. The most recent study of the issue was carried out by \citet{pest07} who used a compilation of all sources detected prior to 2005 \citep{pest05}. Carrying out a statistical analysis of the maser population and modeling the spatial distribution of the masers, the luminosity function was modeled by those authors as a power-law with sharp cutoffs and having an index between --1.5 and --2.

There are two problems facing studies of the luminosity function. First, most studies use a catalog of methanol masers compiled from various surveys, many of them targeted (towards IRAS sources or OH masers), and with different sensitivities. However, unbiased searches have all shown that targeted searches, especially towards IRAS sources, underestimate the number of sources by a factor of 2 or greater \citep{pand07b,szym02,elli96}. For the best estimate of the luminosity function, one should employ a blind survey, since it is possible to analyze the limitations such as completeness reliably. Among the several blind surveys to date, by far the most sensitive one is the Arecibo Methanol Maser Galactic Plane Survey (AMGPS; \citealt{pand07a}). The AMGPS covered an area of 18.2 square degrees between Galactic longitudes of 35\degr~and 54\degr, and detected a total of 86 sources. This survey has a 95\% probability of detection at a peak flux density of 0.27 Jy although sources as weak as 0.13 Jy have been detected. This makes the AMGPS catalog ideal for determining the luminosity function of 6.7 GHz methanol masers, especially at faint luminosities, although the relatively small size of the sample and area covered by the survey results in large statistical uncertainties.

A more formidable problem is the determination of distances to the sources. Measuring distances is an old and challenging problem in observational astrophysics. While trigonometric parallax is the most reliable method for determining distances, it cannot be applied readily to a large sample of sources. The usual technique for Galactic sources is to use a Galactic rotation curve to determine kinematic distances. While kinematic distances can have significant errors at times (e.g. \citealt{xu06}), recent work using Very Large Baseline Interferometry (VLBI) has discovered systematic proper motions in young massive star forming regions, which in principle can be taken into account to improve kinematic distance estimates \citep{reid09}.

A second challenge in the use of kinematic distances arises from an ambiguity between two distances (a ``near'' distance and a ``far'' distance) for sources located within the solar circle in the first and fourth Galactic quadrants. The most popular method to resolve the kinematic distance ambiguity is to measure an absorption spectrum towards the source (which has to have associated continuum radiation), and compare the velocities of the absorption lines with that of the source and the tangent point (see Fig. 1 of \citealt{kolp03} for an illustration of this technique). Distance ambiguities have been successfully resolved towards ultracompact \hii~regions using 21 cm \hi~absorption \citep{kuch94,kolp03,fish03} and 6 cm formaldehyde absorption \citep{aray02,wats03,sewi04}.

\hi~absorption at 21 cm has been used by \citet{pand08} to resolve the distance ambiguity towards 34 6.7 GHz methanol masers that have either directly associated 21 cm continuum, or belong to a cluster harboring a 21 cm continuum source.  However, most methanol masers do not have any detectable radio continuum, presumably due to the young age of the exciting massive young stellar object. Hence, absorption line experiments can be used to resolve distance ambiguities towards only a small sample of 6.7 GHz methanol masers.

\citet{burt78} and \citet{lisz81} discovered that several \hi~self-absorption features in \hi~maps of the Galactic plane correlated with CO emission features, and hence hypothesized that \hi~self-absorption could be used to determine distances to molecular clouds. \hi~self-absorption arises from cold \hi~in the foreground absorbing warmer radiation of the background at the a specific radial velocity within the velocity range covered by the background gas. Hence, only molecular clouds at the near kinematic distance can display \hi~self-absorption, since at the far distance there is no background emission at the radial velocity of the cloud. The theoretical work of \citet{flyn04} shows that molecular clouds have enough opacity in cold \hi~to exhibit self-absorption against strong 21 cm backgrounds (such as in the Galactic plane). \hi~self-absorption has been used by \citet{jack02} and \citet{busf06} to resolve the distance ambiguity towards molecular clouds and massive young stellar candidates.

In this paper, we describe work to resolve the distance ambiguity towards 6.7 GHz methanol masers detected in the AMGPS using \hi~self-absorption. The sources were observed in CO ($J=2-1$) and NH$_3$ to determine their systemic velocities and the line profiles of thermally excited molecular emission. The maser emission of most sources have multiple emission components, which at times are spread over as much as 20~\kms. While the central velocity of the maser emission is usually within a few \kms~of the systemic velocity, it can at times be offset by more than 10 \kms, and is hence not as reliable as velocities derived from thermal molecular emission. With distances determined from \hi~self-absorption, we will then look at the distribution of sources in the Galaxy and derive the methanol maser luminosity function. This in turn will be used to compare the methanol maser population in our Galaxy with that in nearby external galaxies.

\section{Observations and Data Reduction}

The CO ($J=2-1$) observations were made in December 2006 using the 10 m Submillimeter Telescope\footnote{The SMT is operated by the Arizona Radio Observatory, Steward Observatory, University of Arizona, with partial support from the Mt. Cuba Astronomical Foundation.} (SMT) at Mt. Graham operated by the Arizona Radio Observatory. We used the 1.3 mm receiver using ALMA single sideband mixers. The mixers make both sidebands available simultaneously, the sideband rejection being 18--21 dB. We used the setup with the upper sideband including the \coi~$(J=2-1)$ line and the lower sideband including the \coii~$(J=2-1)$ line. For a backend, we used the filterbanks (FFBs) which have 1 GHz bandwidth with 1024 channels yielding a velocity coverage of $\sim$ 1400~\kms~and a velocity resolution of 1.4~\kms. The system temperature ranged from 175~K under good weather conditions to 600~K at low elevations in moderate weather. Since carbon monoxide is very prevalent in the Galaxy, we used absolute position switching, choosing the off-positions from the BU-FCRAO Galactic Ring Survey (GRS; \citealt{jack06}). The on-source integration time was 2 minutes per source. Pointing and focus checks were done on Uranus at regular intervals. 

The data were reduced using the CLASS software (http://www.iram.fr/IRAMFR/GILDAS) and standard procedures. A first order polynomial was used for baselining the spectra. The data were then imported into IDL, and the systemic velocity was determined through a Gaussian fit of the \coii~line using the procedure ``xgaussfit''\footnote{From http://fuse.pha.jhu.edu/analysis/fuse\_idl\_tools.html -- the FUSE IDL tools website.}. We focused on the \coii~spectra since this isotopologue avoids the problems with blended lines from the larger linewidths, and self-absorbed lines that are more common in \coi. Although the velocity resolution is 1.4~\kms, most sources were detected at high enough signal to noise ratio to have their line velocities measured to an accuracy of a fraction of a~\kms. In most cases, the CO line velocity that is within 5~\kms~of the 6.7 GHz methanol emission was chosen as the systemic velocity of the source. 

A few sources displayed two \coii~peaks within 5~\kms~of the maser emission, resulting in an ambiguity in the determination of the systemic velocity of the maser. The CS ($J=5-4$) line was observed in these sources using the APEX telescope\footnote{This publication is based in part on data acquired with the Atacama Pathfinder Experiment (APEX). APEX is a collaboration between the Max-Planck-Institut f\"{u}r Radioastronomie, the European Southern Observatory, and the Onsala Space Observatory.}. The observations were carried out in September 2008 using the APEX-1 single sideband receiver \citep{vass08}. The receiver was tuned to the CS ($J=5-4$) line at 244.936 GHz. The Fast Fourier Transform Spectrometer (FFTS; \citealt{klei06}) was used for the backend with a bandwidth of 1 GHz and 16384 channels, yielding an effective velocity resolution of 0.12~\kms. The observations were made in position switched mode, with an on-source integration time of 3 minutes per source. The pointing was checked every hour, and the focus was verified at the start of the observations. These data were also reduced using the CLASS software. A low order polynomial baseline was subtracted and the center of emission was determined using the CLASS procedure ``gauss'' (``minimize'' in CLASS90).

We observed 15 sources for CS ($J=5-4$) emission, of which 8 sources did not yield any detection to a $1\sigma$ main beam temperature limit of $\sim 0.04$ K (at 0.6 \kms~resolution). For these sources, we use the \nhiii~(1,1) transition observed at the 100 m MPIfR Effelsberg telescope\footnote{Based on observations with the 100-m telescope of the MPIfR (Max-Planck-Institut f\"{u}r Radioastronomie) at Effelsberg.} (Pandian et al., in preparation) to discriminate between the two \coii~velocities. 

\section{Resolving the kinematic distance ambiguity}
\begin{figure}
\epsscale{0.9}\plotone{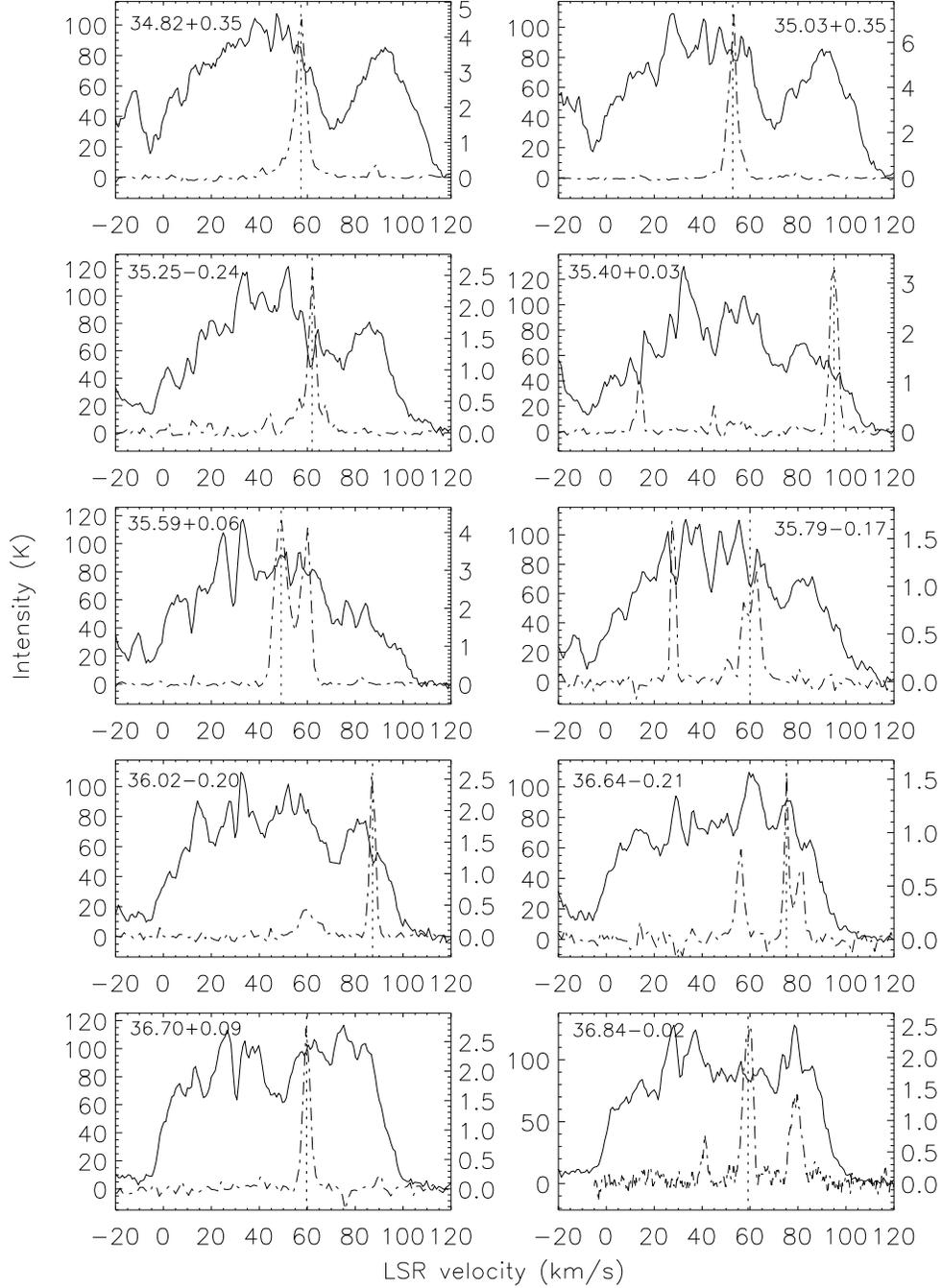}
\caption{Overlays of \hi~and \coii~data. The solid line shows the \hi~spectrum derived from VGPS data at the maser position, while the dashed line shows the \coii~spectrum at the same location. The \coii~data correspond to the $J=2-1$ transition unless otherwise mentioned in the text (\S3.1). The left and right axes indicate the \hi~and \coii~antenna temperatures respectively. The systemic velocity is indicated by the dotted line for clarity. \label{cohi}}
\end{figure}

The kinematic distance ambiguity was resolved by looking for the presence or absence of \hi~self-absorption at the systemic velocity of the source, with the presence of self-absorption implying the near kinematic distance. The \hi~data were taken from the VLA Galactic Plane Survey (VGPS; \citealt{stil06}), which has a resolution of $1\arcmin \times 1\arcmin \times 1.56$\kms. Some sources are also close to compact \hii~regions which have had their distance ambiguity resolved using previous absorption line studies. In these cases, we verified the classification using \hi~self-absorption, and adopted the results from previous work in cases of ambiguous \hi~line profiles. Attention was paid to the line profiles of \hi~and \coii, which should be similar in cases of self-absorption. We also allowed for offsets up to $\sim 1$ \kms~between the CO and \hi~lines when considering the presence or absence of self-absorption. The resulting distances and isotropic luminosities of the sources are shown in Table 1, and selected individual sources are discussed in section \ref{indisources}. We show overlays of the \hi~and \coii~spectra in Fig.~\ref{cohi}.

The distances in Table 1 are calculated using the modified kinematic distance technique outlined in Reid et al. (2009), which was found to minimize the error between trigonometric parallax distances and kinematic distances towards an ensemble of massive star forming regions. First, systemic velocities were calculated in a modified LSR frame of reference with $(U_0,~V_0,~W_0) = $ (10.0,~5.25,~7.17) for the solar motion. Kinematic distances were then calculated using a flat rotation curve with $R_0 = 8.4$ kpc, and $\Theta_0 = 254$~\kms, with the assumption that massive star forming regions rotate 15~\kms~slower than the circular rotation speed. Uncertainties were calculated by including a 7~\kms~uncertainty in the modified LSR velocity.

\subsection{Notes on selected sources}\label{indisources}
{\noindent \it 34.82+0.35} -- This source is close to the G34.4 massive star forming region (Galactic longitude and latitude coordinates $\sim 34.4\degr,~0.23\degr$). Its systemic velocity of 57.4 \kms, which is close to that of G34.4 (57 \kms), suggests that the two regions are physically associated. The \hi~spectrum at the source location does not show self-absorption, although nearby locations in the same cloud complex show prominent self-absorption. Since the G34.4 complex is thought to be at the near kinematic distance, this maser source is also tentatively placed at the near distance.

{\noindent \it 35.03+0.35} -- The \hi~line profile in this source is ambiguous with an emission line inside an absorption feature. \citet{wats03} determine this source to be at the far distance using 6 cm formaldehyde absorption. Hence, this source is classified at the far distance.

{\noindent \it 35.39+0.02 \& 35.40+0.03} -- These two sources are part of a cluster with the same CO velocity. There is weak \hi~self-absorption in both sources (only 35.40+0.03 is shown in Fig.~1) and hence both sources are classified to be at the near distance.

{\noindent \it 35.59+0.06} -- There are two \coii~lines in this source at 49.1 and 59.3~\kms. APEX CS observations show the 49.1 \kms~component to be associated with the methanol maser. The lack of \hi~self-absorption at this velocity places this source at the far distance. This is confirmed by a similar classification by \citet{kolp03} for a related source using \hi~absorption.

{\noindent \it 35.79--0.17} -- There are two \coii~lines in this source at 57.5 and 62.4~\kms. There is CS emission at 62.5~\kms, while the \nhiii~line peaks at 61.3~\kms. Thus the \coii~line profile arises from the presence of different cores in the molecular cloud rather than self-absorption. The presence of \hi~self-absorption near the CO velocity places this source at the near distance.

{\noindent \it 36.64--0.21} -- There are two \coii~lines at 75.2 and 80.8~\kms. The \hi~spectra show the 75.2~\kms~component to be at the far distance, while the 80.8~\kms~component is at the near distance. The two features are also seen to be distinct clouds from the GRS datacube which shows the 80.8~\kms~component to have a large angular extent, in contrast to the 75.2~\kms~component which is extremely compact. The \nhiii~line shows emission at 74.8~\kms~indicating that the maser source is at the far distance.

{\noindent \it 36.84--0.02} -- The CO ($J=2-1$) spectrum was taken at an incorrect location on the sky, and hence the GRS \coii~($J=1-0$) spectrum is shown in Fig.~1 (the systemic velocity being verified by APEX CS observations). The presence of \hi~self-absorption places this source at the near distance. The GRS datacube also shows this source to be distinct from the source 36.70+0.09 even though they are at very similar velocities.

{\noindent \it 37.02--0.03 \& 37.04--0.04} -- The source 37.04--0.04 (J2000 coordinates: 18$^h$~59$^m$~4$^2$.41, 3\degr~38\arcmin~32\arcsec.8; peak flux density = 4.13 Jy) is a new detection discovered in MERLIN observations to determine accurate positions of the AMGPS sources (Pandian et al., in preparation). The GRS datacube shows both sources to be part of a molecular cloud that spans $\sim 0.4$\degr~in Galactic longitude. The presence of \hi~self-absorption suggests that this cloud (and both maser sources) are at the near kinematic distance (only 37.02--0.03 is shown in Fig.~1).

{\noindent \it 37.38--0.09} -- There are two \coii~lines in this source at 57.7 and 69.2 \kms. \nhiii~observations at Effelsberg show the 57.7 \kms~component to be associated with the maser. The lack of \hi~self-absorption shows that this source is likely to be at the far distance. This agrees with the distance determination of the nearby \hii~region 37.37--0.07 by \citet{kuch94} based on \hi~absorption (towards continuum in the \hii~region).

{\noindent \it 37.53--0.11} -- The \hi~spectrum in this source shows several features with intensities below 0~K. Interpreting these as absorption lines towards continuum emission, this source is at the far distance since absorption lines are seen between the systemic velocity of the source and the tangent point velocity (84 \kms).

{\noindent \it 37.55+0.19} -- The \coii~line profile suggests that this line is self-absorbed, which is confirmed by Effelsberg \nhiii~observations. Taking this into account, there is prominent \hi~self-absorption in this source, and hence we classify it at the near kinematic distance.

{\noindent \it 37.60+0.42} -- The \hi~spectrum shows absorption at the CO velocity, though the line profiles do not match well. The discrepancy however is not severe such as in 38.92--0.36 (below). Hence, we tentatively classify this source to be at the near distance.

{\noindent \it 37.77--0.22} -- The \hi~spectrum in this source is similar to that in 37.53--0.11. The presence of absorption lines at velocities between the source velocity and that of the tangent point indicates that this source is at the far distance.

{\noindent \it 38.26--0.08} -- There are two \coii~lines at 11.6 and 17.7 \kms, the former being associated with the maser as deduced from \nhiii~observations. The \hi~spectrum shows an absorption feature that is misaligned with the CO velocity by almost 2 \kms~making it unlikely to be self-absorption. We thus place this source at the far distance.

{\noindent \it 38.92--0.36} -- The \hi~spectrum shows a very broad absorption feature around the CO velocity. Since the feature is too broad to be interpreted as self-absorption from cold \hi, and the CO and \hi~line profiles are different, we locate this source to be at the far distance, and attribute the \hi~feature to the intrinsic line profile rather than absorption.

{\noindent \it 39.54--0.38} -- This is another ambiguous case with the \hi~spectrum showing a broad absorption feature that does not match the line profile of both \coii~($J=2-1$) and the ($J=1-0$) lines. Inspection of the GRS datacube suggests that this source is distinct from 39.39--0.14 that is at similar LSR velocity (and which shows prominent \hi~self-absorption). Moreover, examining the \hi~and \coii~($J=1-0$) spectra at other locations in the molecular cloud shows no self-absorption. Hence, this source is tentatively classified to be at the far distance.

{\noindent \it 41.08-0.13} -- The \hi~spectrum shows features that look like self-absorption at some locations in the molecular cloud, while other locations including the maser position show no self-absorption. Thus, this is another ambiguous case whose distance ambiguity cannot be reliably resolved using \hi~self-absorption. Taking into account the results of the nearby sources at similar LSR velocities, 41.12--0.22 (the distance ambiguity for which has been resolved by previous absorption line studies), 41.16--0.20 and 41.23--0.20, we place this source at the far distance though with low confidence. 

{\noindent \it 41.12--0.11} -- The \coii~spectrum in this source is complex with multiple emission components. Effelsberg \nhiii~data and APEX CS data show the 38.1 \kms~component (which is also the strongest) to be associated with the maser. This \hi~spectrum shows absorption that is misaligned relative to the CO emission by more than 2 \kms. Hence, we classify this source to be at the far distance.

{\noindent \it 41.87-0.10} -- Since the coordinates of this source are uncertain (as explained in \citealt{pand07a}), we use the coordinates of the nearest 24 $\mu$m point source in the MIPSGAL survey \citep{care09}, and also examine other locations in the parent molecular cloud. As in source 39.54--0.38, there is a broad HI absorption feature whose line profile does not correspond that of the relatively narrow CO line. Hence, this source is also classified to be at the far distance.

{\noindent \it 42.30-0.30} -- The \hi~spectrum shows an absorption feature that is offset relative to the CO emission by $\sim 2$ \kms~(similar to 38.26--0.08). Hence, this source is placed at the far distance.

{\noindent \it W49N region} -- There are 5 methanol masers in the W49N region: 43.15+0.02, 43.16+0.02, 43.17+0.01, 43.17-0.00, and 43.18-0.01. The CO line shows emission at two velocities around 0 and 10 \kms~respectively. All sources in this region are placed at a distance of $11.4 \pm 1.2$~kpc based on proper motion studies of H$_2$O masers by \citet{gwin92}. This is also in very close agreement with the kinematic far distance (11.4 -- 12.0 kpc) of the sources.

{\noindent \it 43.80-0.13} -- There is a \hi~absorption feature in the vicinity of CO emission although there is poor correspondence between the \hi~and CO line profiles. Absorption line studies of \citet{wats03} find this source to be at the far distance. This highlights the importance of the similar line profiles in the \hi~and CO spectra for the \hi~feature to be described as self-absorption.

{\noindent \it 44.64-0.52} -- This \hi~spectrum in this source is similar to that in 38.92--0.36, showing a broad ``absorption'' feature and a narrow CO line, making it difficult to interpret the former as self-absorption. Hence, this source is classified to be at the far distance.

{\noindent \it W51 complex} -- There are 10 methanol masers in the AMGPS (48.89--0.17, 48.90--0.27, 48.99--0.30, 49.35+0.41, 49.47--0.37, 49.48--0.40, 49.49--0.37, 49.49--0.39, 49.60--0.25 and 49.62--0.36) that can be attributed to the W51 molecular cloud complex. Among these, 48.90--0.27, 48.99-0.30, and 49.35+0.41 belong to the 68 \kms~component \citep{carp98} while the remaining sources are in the W51 giant molecular cloud. Using the trigonometric parallax distance of W51~IRS2 by \citet{xu09}, these sources are placed at a distance of $5.1^{+2.9}_{-1.4}$~kpc. This is also close to the kinematic tangent point distance (5.5 kpc) of the sources.

{\noindent \it 52.92+0.41} -- The LSR velocity of this source is greater than the tangent point velocity. Hence, this source is classified to be at the tangent point.

{\noindent \it 53.14+0.07 \& 53.62+0.04} -- These two sources are part of a cloud complex that spans a range of $\sim 1.2$\degr~in Galactic longitude. Both sources show prominent \hi~self-absorption and are hence at the near distance.

\section{Discussion}
\subsection{Comparing maser and systemic velocities}
\begin{figure}
\epsscale{0.7}\plotone{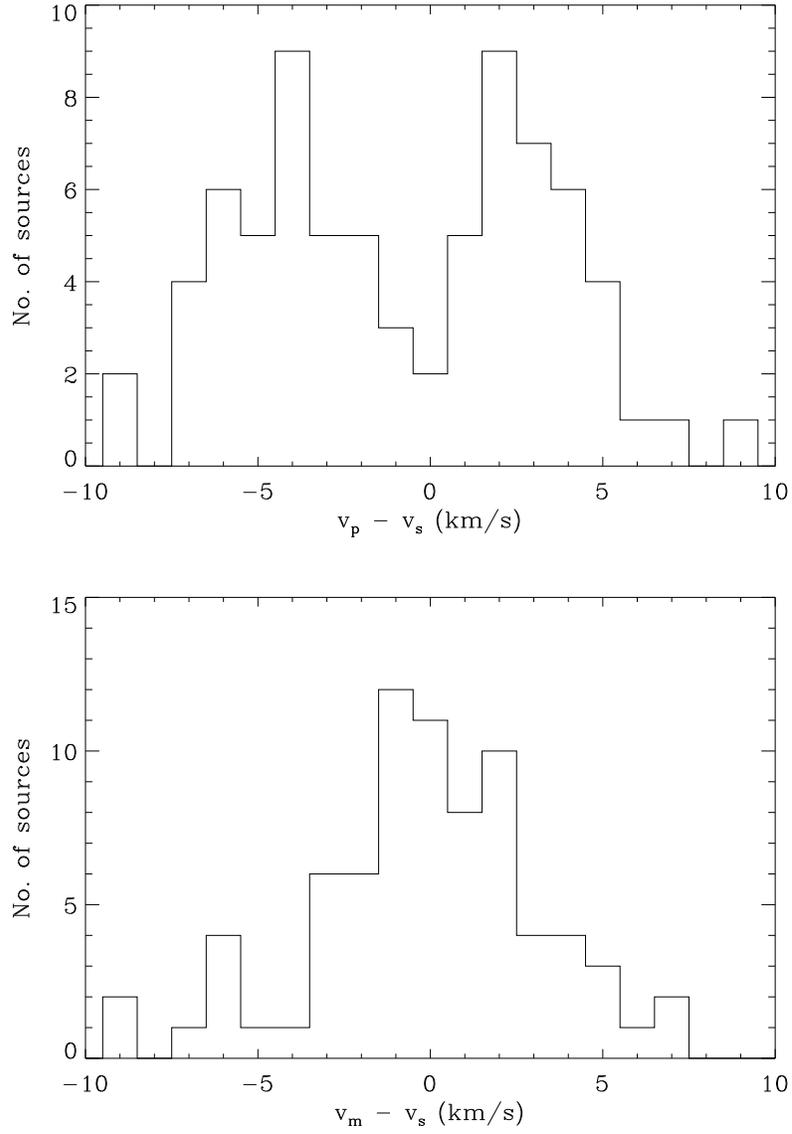}
\caption{Histograms of the difference between the peak velocity of maser emission and the systemic velocity (top panel) and that of the mean velocity of maser emission and the systemic velocity (bottom panel). \label{vdiff}}
\end{figure}

Fig.~\ref{vdiff} shows the histogram of the difference between (a) the peak velocity of maser emission ($v_p$) and the systemic velocity ($v_s$) and (b) the central velocity of maser emission ($v_c$, the arithmetic average of the minimum and maximum velocities of maser emission; all quantities are shown in Table 1). The sources in W49N and W51 main are not included in this plot since the systemic velocities of these sources are not well known. Fig.~\ref{vdiff} shows a puzzling double peaked histogram for the difference between the maser peak velocity and the systemic velocity. On the other hand, the central maser velocity has a much better agreement with the systemic velocity though with large scatter. The average value of $v_c - v_s$ is $0.08 \pm 3.8$ \kms. While this average value is similar to that obtained by \citet{szym07}, the scatter in the AMGPS sample is much larger. The fraction of sources where the difference between $v_c$ and $v_s$ is less than 3 \kms~is also smaller in our sample -- 68\% versus 83\% seen by \citet{szym07} in their work. Nevertheless, Fig.~\ref{vdiff} shows that in the absence of molecular line data, the central maser velocity is a better estimate of the systemic velocity than the peak emission velocity.

\subsection{Distribution in the Galaxy}

\subsubsection{Galactocentric distance}

\begin{figure}
\plotone{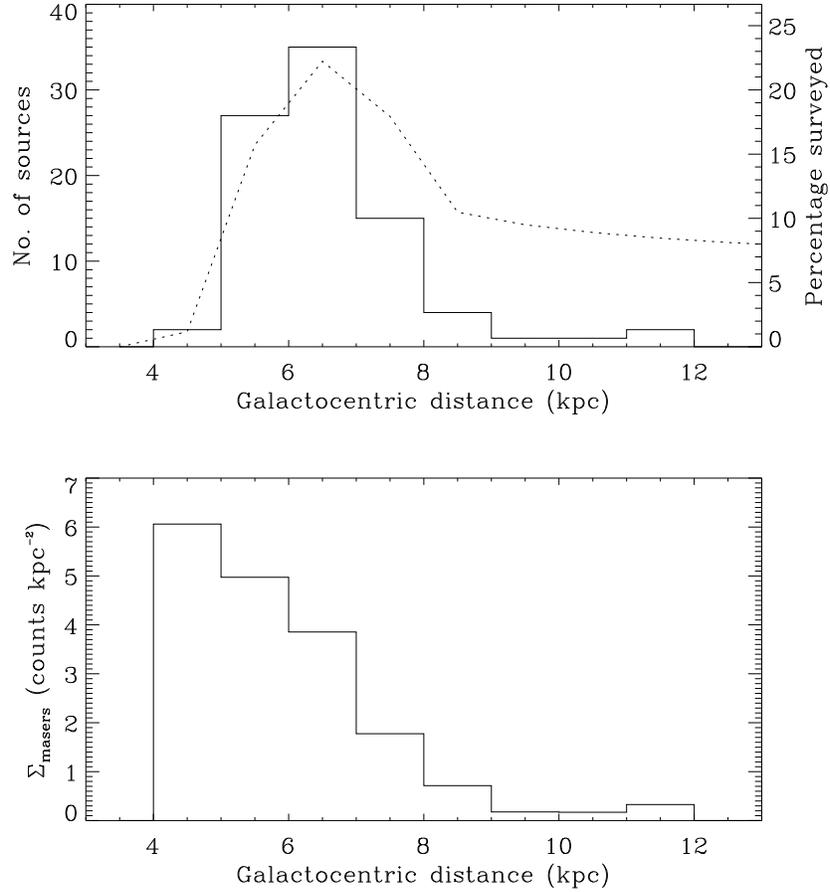}
\caption{The top panel shows the histogram of Galactocentric distances to the methanol masers in AMGPS. The dotted line (right ordinate) shows the fraction of area in the Galaxy of annuli represented by each distance bin that was covered by the survey. The bottom panel shows the derived surface density of methanol masers in the Galaxy. \label{rhist}}
\end{figure}

The top panel of Fig.~\ref{rhist} shows the distribution of Galactocentric distances for the AMGPS sample. It should be kept in mind that any survey that does not cover the entire Galaxy will have a selection function in terms of sampling the area/volume at different Galactocentric radii. The dotted line in the top panel of Fig.~\ref{rhist} shows an estimate of this selection function by calculating the fraction of the total area in each distance bin annulus that was within the survey limits ($35\degr \la l \la 54\degr$, where $l$ is the Galactic longitude). A more unbiased description of this distribution is presented in the bottom panel of Fig.~\ref{rhist} where the surface density (in counts kpc$^{-2}$) is shown as a function of Galactocentric distance. It should be noted that the survey limit of $l=35\degr$ translates to a minimum Galactocentric distance of 4.8 kpc, and so the surface density shown in the 4--5 kpc bin has a large uncertainty. The shape of the distribution is in very good agreement with that of \citet{pest07} although the absolute numbers are higher by almost a factor of 3. Considering that the distribution of \citet{pest07} is derived from a compilation of sources detected from a number of surveys (both targeted and unbiased) with different sensitivities, the close agreement of the shape of the distribution is remarkable. The difference in absolute numbers arises from both the shallower depth of previous surveys, and the fact that the entire Galaxy is not covered by the surveys in the compilation (leading to the surface density being underestimated by \citealt{pest07}).

\subsubsection{Locations in the Milky Way}

\begin{figure}
\plotone{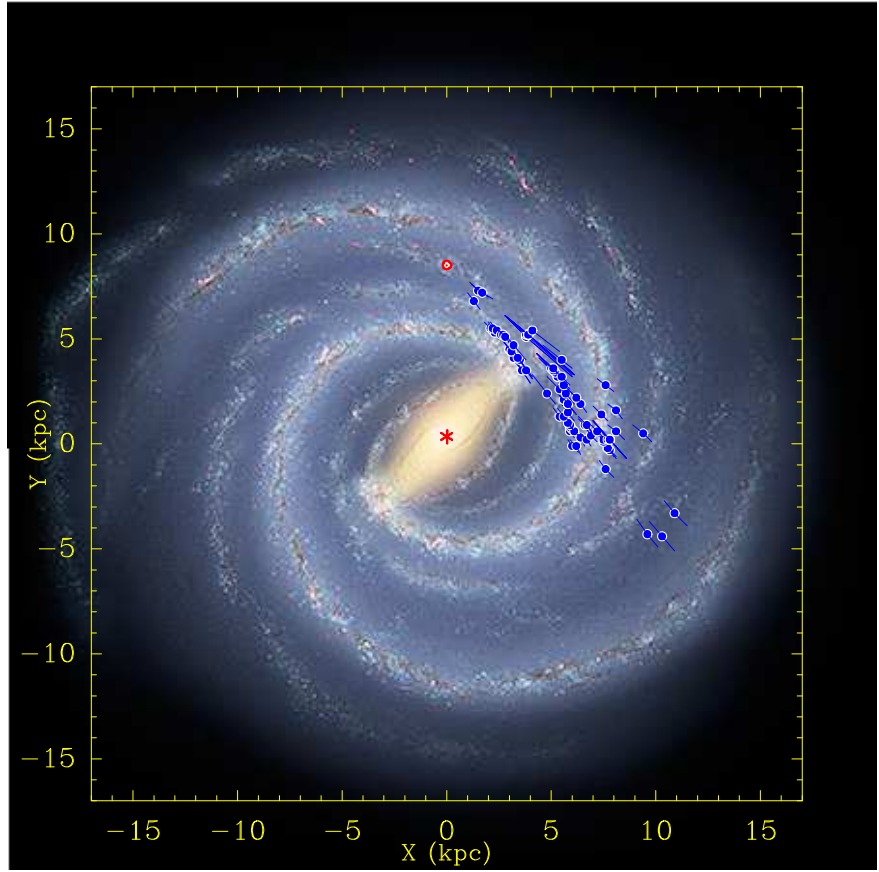}
\caption{Locations of the methanol masers (blue dots with lines indicating the uncertainties) discovered in AMPGS. The Galactic center (red asterisk) and the Sun (red solar symbol) are located at (0, 0) and (0, 8.5) respectively. The background is an artist's conception of the Milky Way (R. Hurt: NASA/JPL-Caltech/SSC). The distances in Table 1 are scaled to an $R_0$ of 8.5 kpc in this illustration. \label{mw}}
\end{figure}

Fig.~\ref{mw} shows the locations of the AMGPS methanol masers in a face-on view of the Galaxy superposed on an artist's conception of the Milky Way. There is a significant cluster of sources near the Crux-Scutum arm (close to the Galactic bar). Taking the error bars into account, it is possible to reconcile the positions of most sources with those of the Crux-Scutum, Carina-Sagittarius and Perseus spiral arms.  The sources 53.14+0.07 and 53.62+0.04 are located in the Local arm, and 49.41+0.33 appears to be in the Outer arm. The three distant sources, 36.92+0.48, 38.66+0.08 and 42.70-0.15 lie in between the Outer arm and the Crux-Scutum arm, and may trace a spur from the Outer arm. Since most distances to the sources are kinematic and have significant uncertainties associated with them (even though an estimate of some of the proper motions are taken into account by using the modified technique of \citealt{reid09}), it is not possible to carry out any quantitative analysis of the spiral arms traced by AMGPS methanol masers.

\subsubsection{Vertical Distribution}

\begin{figure}
\plotone{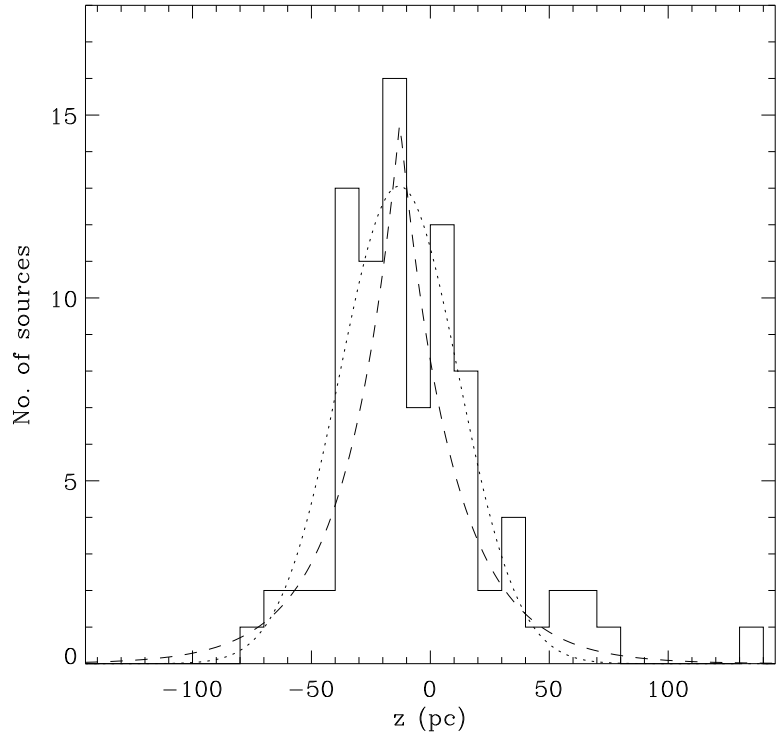}
\caption{Vertical distribution of the AMGPS sources. The dotted and dashed lines show Gaussian and exponential fits to the distribution respectively. The scale height derived from the fits are 30 pc and 20 pc from the Gaussian and exponential fits respectively. \label{zdist}}
\end{figure}

Fig.~\ref{zdist} shows the vertical distribution of the AMGPS sources about the Galactic plane. Also shown in Fig.~\ref{zdist} are Gaussian and exponential fits to the observed distribution. Both distributions peak at negative $z$, reflecting the fact that the Sun is located above the Galactic plane as defined by the IAU. The locations of the peak (13.1 pc and 12.8 pc for Gaussian and exponential fits respectively) are very close to the estimated position of the Sun above the Galactic plane ($\sim 16$ pc; \citealt{hamm95}). The half-width at half maximum of the Gaussian fit is $\sim 30$ pc, while the scale height of the exponential fit is $\sim 20$ pc. These numbers are significantly less than the scale height of the Galactic thin disk, which is thought to be around 100 pc. This may in part be a selection effect caused by the relatively small coverage of AMGPS in terms of Galactic latitude ($\sim \pm 0.4\degr$). However, a very similar scale height has been observed by \citet{van96} using methanol masers detected prior to 1996. 

\begin{figure}
\plotone{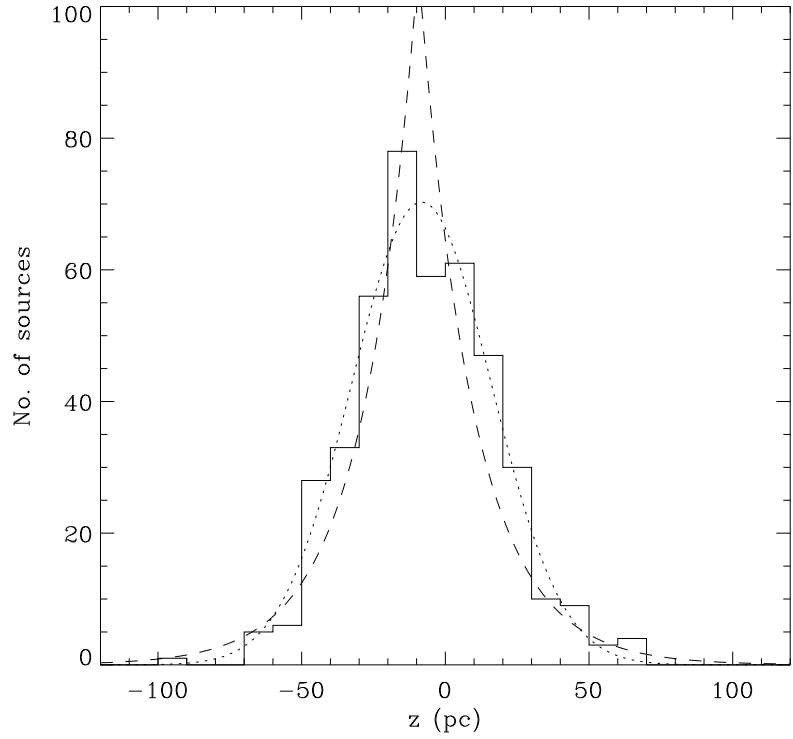}
\caption{The vertical distribution of Infrared Dark Clouds. As in Fig. \ref{zdist}, dotted and dashed lines show Gaussian and exponential fits to the distribution respectively. The scale heights derived for the infrared dark clouds are identical to that of 6.7 GHz methanol masers in Fig. \ref{zdist}. \label{irdcdist}}
\end{figure}

To investigate this issue further, we looked at the vertical distribution of infrared dark clouds (IRDCs), which are thought to be sites of imminent and ongoing massive star formation. Since the clouds are seen as extinction features against the Galactic mid-infrared background, they are thought to be at the near kinematic distance. Using the catalogs of \citet{simo06} and \citet{jack08}, which cover the first and fourth Galactic quadrants respectively, we determined the vertical distribution of IRDCs (Fig. \ref{irdcdist}). The scale heights of Gaussian and exponential fits are essentially identical to what is seen for 6.7 GHz methanol masers. Moreover, the scale height of compact \hii~regions in the Galaxy is observed to be around 35 pc (\citealt{fish03} and references therein). This strongly suggests that the low scale heights observed for 6.7 GHz methanol masers using the AMGPS sample is real, and reflects the scale height of newly born massive stars in the Galaxy.

\subsubsection{Near/Far Asymmetry}

Of the 87 sources in the AMGPS, 21 are classified to be at the near distance, and one source is placed at the tangent point. We can estimate whether this is realistic by calculating the fraction of the volume sampled in the survey that is in between the Sun and the tangent point (and thus corresponds to the near kinematic distance). To get meaningful results, this must be weighted by some function that takes into account the lower density of sources at large Galactocentric distances. Using the surface density of Fig.~\ref{rhist} as the weighting function, 28\% of the sources should be at the near distance. This is very close to the 24\% observed in this study. In contrast, \citet{sobo05} suggest confinement of kinematic distances to the near kinematic distance to produce realistic statistical results. Our work coupled with the results of \citet{kolp03}, who observed $\sim 25\%$ of \hii~regions to be at the near distance, indicates that the assumption of \citet{sobo05} that methanol masers are statistically likely to be at the near distance is flawed, and may be due to the fact that the masers considered by them were found in much shallower surveys than ours. Deep surveys, such as the AMGPS, are sensitive to most methanol masers in the Galaxy, which is reflected in the close agreement of the observed near/far asymmetry with that expected from volume considerations.

A curious feature that can be observed in Table 1 is that all sources between Galactic longitudes of 40.3\degr~and 53\degr~are at the far distance. This may indicate a dearth of young massive star forming regions in the near side of the Carina-Sagittarius spiral arm between Galactic longitudes of 35\degr~and 49\degr~(corresponding to Galactocentric longitudes, $\beta$, of 15\degr~and 45\degr~in the terminology of an ideal log-periodic spiral arm, where $\beta$ is measured clockwise from the line joining the Sun and Galactic center). Fewer than five sources in the AMGPS catalog have systemic velocities close to what is expected from this region ($\sim 25$ \kms~at $l=35\degr$ to $\sim 55$ \kms~at $l=49\degr$). A similar paucity is seen in the IRDC catalog of \citet{simo06}, though there are a larger number of sources (in the radial velocity range given above) than in the AMGPS catalog. Accurate distance measurements to the candidate sources in the region, and future work with other star formation indicators is required to verify this observation.

\subsection{Luminosity function of 6.7 GHz methanol masers}

Once the kinematic distance ambiguity is resolved, one can calculate the isotropic luminosities of the masers (using their integrated fluxes) and the methanol maser luminosity function. The luminosities range from $4.0 \times 10^{-9}~L_\sun$ to $2.0 \times 10^{-4}~L_\sun$. The mean luminosity (computed logarithmically) is $8.5 \times 10^{-7}~L_\sun$, while the median luminosity is $9.1 \times 10^{-7}~L_\sun$. 

\begin{figure}
\plotone{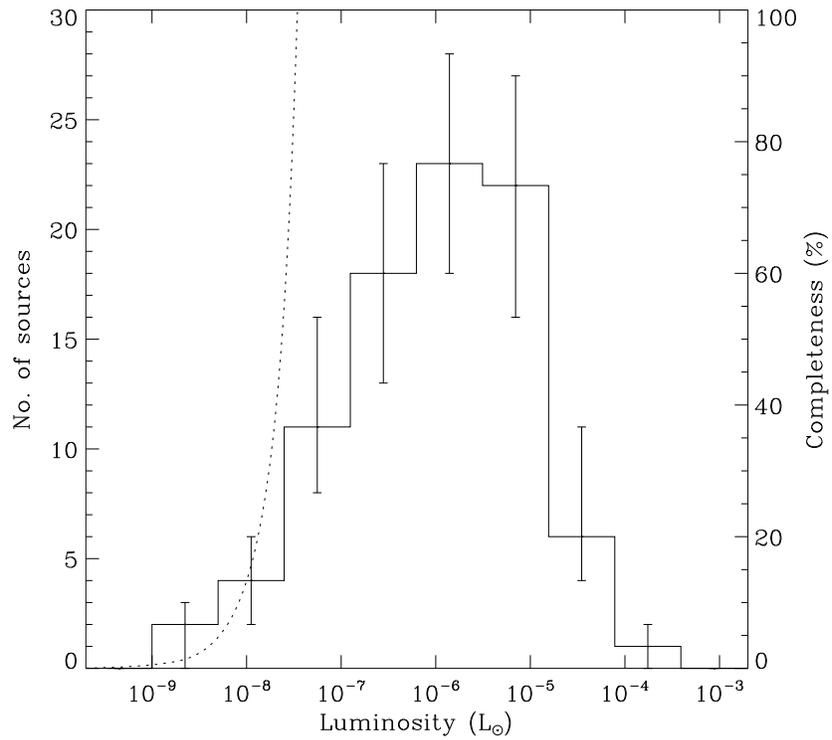}
\caption{The luminosity function of 6.7 GHz methanol masers. The error bars shown include uncertainties in the kinematic distance, uncertainty in the resolution of kinematic distance ambiguity using \hi~self-absorption, and statistical uncertainties. The dotted line shows the completeness (right hand scale) of the AMGPS survey as a function of luminosity, the survey being essentially 100\% complete for $L > 3.3 \times 10^{-8}~L_\sun$. \label{lfunction}}
\end{figure}

Fig. \ref{lfunction} shows the luminosity function of 6.7 GHz methanol masers. Errors in the luminosity function may arise from three sources: (i) formal uncertainties in the calculated luminosities (which arise from uncertainties in the kinematic distance), (ii) the success rate of the \hi~self-absorption technique to resolve the kinematic distance ambiguity, and (iii) statistical uncertainties from the finite sample. \citet{busf06} carried out a confidence test of the \hi~self-absorption technique by randomly shifting the CO spectrum in velocity and looking for self-absorption on the shifted data. Using this test, they estimate the success rate in determining the correct solution for the distance ambiguity to be $\sim 80\%$. Hence, we carried out simulations in which 20\% of the sources which had the distance ambiguity resolved solely from \hi~self-absorption had their resolution (near/far) reversed followed by recalculation of the luminosity function. The error bars indicated in Fig.~\ref{lfunction} reflect the results of these simulations in addition to taking into account the formal uncertainties in the luminosities themselves, as well as the statistical uncertainties.

To interpret Fig.~\ref{lfunction}, we have to estimate the completeness of the survey for various luminosities. Fig.~3 of \citet{pand07a} shows the completeness of the AMGPS as a function of peak flux density. However, the detection probability of the matched filtering algorithm used for source detection will be a function of the integrated flux. Hence, we repeated the simulations described in \citet{pand07a} for different linewidths, and determined that the survey is complete for an integrated flux of 0.08 Jy~\kms. We can now calculate the limiting distance at which all sources at a given luminosity are detected. The fraction of the area defined by the limiting distance and the survey limits to the total area covered by the survey gives an estimate of the completeness at this luminosity. However, the surface density of methanol masers decreases with increasing Galactocentric radius as shown in Fig.~\ref{rhist}. Hence, the area should be weighted by the surface density to obtain an accurate estimate of the completeness. The dashed line in Fig.~\ref{lfunction} shows the result of this calculation, indicating that all sources in the survey region with luminosities greater than $3.5 \times 10^{-8}~L_\sun$ are detected by AMGPS.

There are two caveats in the above discussion that should be noted. First, methanol masers are variable. However, significant variability (e.g. flaring behavior) is limited to a small number of sources. Since at any given point of time, one can expect some masers to be in a higher than normal state, while other masers would be in a lower than normal state, variability will not systematically affect the entire survey. The overall shape of the luminosity function would then be relatively unaffected when considering a large sample. Second, the luminosities calculated assume that maser emission is isotropic. On the other hand, maser theory shows that masers are beamed, with the nature and extent of beaming being dependent on the geometry and whether the maser is saturated or not (e.g. \citealt{elit92}). However, little is known about maser geometry or beaming in astrophysical sources. Theoretical models of methanol masers (e.g. \citealt{crag05}) assume a beaming factor, defined as the ratio of optical depths in radial and tangential directions (resulting from an elongated geometry along the line of sight), of 10. The VLBI observations of \citet{mini02} resulted in the detection of core/halo structures, which could mostly be explained by saturated masers in a spherical geometry, although modeling implied large errors on the size of the halos. A spherical geometry however would not have any preferred direction for the beaming, and hence isotropic luminosities would still be accurate. Since the geometries of astrophysical masers are largely unknown, we cannot realistically address the effect of beaming on the luminosity function.

Fig.~\ref{lfunction} shows that the luminosity function rises sharply up to $\sim 10^{-6}~L_\sun$ and declines at higher luminosities. Only the first two bins are severely affected by incompleteness (overall completeness $\sim 95\%$ for the third bin), and hence the continued decline in the number of sources for luminosities below $10^{-7}~L_\sun$ appears to be real. At the high end of the luminosity function, we are limited by the relatively small coverage of the Galaxy by AMGPS. Since sources that have luminosities greater than $10^{-4}~L_\sun$ are rare, one needs wide-area surveys to obtain an accurate census of such objects in the Milky Way. In any case, it is clear that the luminosity function cannot be described by a power law as was suggested in previous work. The behavior of the luminosity function for luminosities below $\sim 10^{-8}~L_\sun$ cannot be constrained by the AMGPS, or any other available sample.

The luminosity function of 6.7 GHz methanol masers also appears to be different from that of OH masers as determined by \citet{casw87}, who found the number of OH masers to increase with decreasing luminosity down to 3 Jy~kpc$^2$ (assuming that the Sun is located 10 kpc from the Galactic center). It should be noted that the unit Jy~kpc$^2$ does not describe luminosity since the linewidth is not taken into account, and most OH masers, like Class-II methanol masers, display multiple emission components. Moreover, the kinematic distance ambiguity is not resolved, and the distinction between the two distances is based on the expected number of sources at near and far distances from volume arguments. It is thus possible that the OH maser luminosity function is different from that determined by \citet{casw87}. However, constructing a figure similar to that of \citet{casw87} using only the peak flux densities of the AMGPS sources results in a shape similar to that of Fig.~\ref{lfunction} with a peak around 500~Jy~kpc$^2$ with a decreasing number of sources at lower $S_pd^2$ (where $S_p$ is the peak flux density and $d$ is the distance). Hence, unless there are significant systematic errors in the distances to the sources, the luminosity functions of 6.7 GHz methanol masers and OH masers are dissimilar.

The ongoing Methanol Multi-Beam survey (MMB; \citealt{gree09}) when completed should cover the entire Galactic Plane but at poorer sensitivity. The 90\% completeness of the MMB is estimated to be between 0.9 and 1.1 Jy \citep{gree09}, which translates to an integrated flux of 0.34--0.41~Jy~\kms~(assuming a linewidth of 0.35~\kms). A crude scaling from the sensitivity of the AMGPS would imply that the MMB will be complete to luminosities around $(1.5-1.8) \times 10^{-7}~L_\sun$, and will thus be able to verify the turn-over of the luminosity function. It is not clear whether the MMB will be able to probe the decline in luminosity function at lower luminosities, though it will determine the high-end very well.

\subsection{Extragalactic methanol masers}

We can use the luminosity function of methanol masers in our Galaxy to revisit the problem of dearth of such sources in external galaxies. To date, there have been searches for 6.7 GHz methanol masers towards the Large Magellanic Cloud (LMC; \citealt{sinc92,elli94,beas96,gree08}), the Small Magellanic cloud (SMC; \citealt{elli94,beas96,gree08}) and M33 \citep{gold08}. Four methanol masers have been discovered in the LMC, but there have been no detections in the SMC or M33. The $3\sigma$ sensitivity limit of the LMC and SMC surveys of \citet{gree08} are 0.27 Jy and 0.4 Jy respectively. In M33, the 3 sigma limit for an individual giant molecular cloud (GMC) is 4 mJy, while the limit on the overall mean emission by combining data for 14 GMCs is 1~mJy \citep{gold08}. The distances adopted for LMC, SMC and M33 are 50 kpc, 60 kpc and 730 kpc respectively \citep{feas99,walk99,brunt05}.

It must be borne in mind that while the luminosity function involves the total integrated flux, surveys are sensitive to the peak flux density of a source. This is especially a factor for strong methanol masers which have a number of spectral features. Among the four 6.7 GHz methanol masers in the LMC, two have a single spectral feature, while two have a more complex spectrum dominated by 3 features, two of which have similar strength \citep{gree08}. There are two approaches that one can use to evaluate the under-abundance of 6.7 GHz methanol masers in external galaxies. The first is to use the peak flux densities of the AMGPS sources, scale them to the distance of each galaxy, and determine the number of sources that should have been detected in the surveys to date. A second approach would be to estimate a mean value of the integrated flux to peak flux density ratio, and use the luminosity function to estimate the number of sources above the threshold corresponding to the survey limits. In the AMGPS, the mean value of this ratio is close to unity ($\sim 1.3$ for very strong sources, and $\sim 0.6$ for weak sources). The results of the two approaches are tabulated under ``Method-I'' and ``Method-II'' respectively in Table 2. The number of methanol masers in the Galaxy has been assumed to be $\sim 1300$ \citep{pand07b,van05}.

An additional factor to consider in this discussion is that the masses and star formation rates in the galaxies in question are different from that of the Milky Way, which will influence the total number of methanol masers in the galaxies. Using the estimates of \citet{isra80}, the star formation rate in the LMC, SMC and M33 are 0.1, 0.02 and 0.2 times that of the Galaxy. If we assume that the number of young massive star forming regions is proportional to the current star formation rate (with an appropriate scaling for the total population of 6.7 GHz methanol masers), then the total number of detectable methanol masers in the LMC, SMC and M33 reduce to 18--20, 2, and 14--18 respectively. The non-detection of methanol masers in the SMC is then not inconsistent with a methanol maser population with similar properties as that of the Galactic sources, though the much lower metallicity casts doubt on this assumption. The LMC population is about a factor of 4--5 under-abundant compared to that in our Galaxy, which is consistent with the estimates of \citet{gree08}. The M33 population is under-abundant by more than a factor of 14 although there is some uncertainty in the latter since only 14 GMCs were sampled in M33. 

This work thus confirms previous results that the methanol maser population in the LMC and M33 are different from that in our Galaxy. For a detailed discussion on the possible reasons for the dearth of methanol masers in these galaxies, we refer the reader to \citet{gree08} and \citet{gold08}. However, the greater under-abundance in M33 in spite of its higher metallicity compared to the LMC is a mystery.


\section{Conclusions}

We used \coii~($J=2-1$) observations and VGPS \hi~data to resolve the kinematic distance ambiguity towards 6.7 GHz methanol masers discovered in the AMGPS. The distribution of the surface density of methanol masers as a function of Galactocentric distance agrees very well with existing estimates in the literature, although we observe the absolute numbers to be more than a factor of 3 higher. The vertical distribution of the sources has a scale height that is $\sim 3$--5 times lower than that of the Galactic thin disk, presumably reflecting the smaller scale height of newly born massive stars. The resolution of the distance ambiguity allowed us to construct a reliable estimate of the luminosity function. Its shape does not agree with that of a power law, but has a peak around $\sim 10^{-6}~L_\sun$ followed by a decline towards lower luminosities. The luminosity function of 6.7~GHz methanol masers also appears to be different from that of mainline OH masers. Using the luminosity function, we derive estimates for the abundance of methanol masers in the LMC, SMC and M33 compared to the Milky Way. We find the under-abundance in M33 to be a factor of 3 higher than that in the LMC in spite of its higher metallicity. Finally, the distribution of sources between near and far distances closely follows the respective volumes sampled by the survey, thus indicating that the assumption of the near kinematic distance on a statistical basis should be avoided.

\acknowledgements
We are very grateful to Kiriaki Xiluri, Gene Lauria, and the operators at the SMT for assistance in obtaining the \coii~spectra for our sample. We thank A. Kerr and the National Radio Astronomy Observatory for use of the ALMA Band 6 mixers at the SMT. We are also grateful to Arnaud Belloche for assistance with the APEX observations, and to Alex Kraus and the operators at Effelsberg for assistance with the \nhiii~observations. We also thank the anonymous referee for useful comments. This work was supported in part by the Jet Propulsion Laboratory, California Institute of Technology, under a contract with the National Aeronautics and Space Administration. This research has made use of NASA's Astrophysics Data System.

\begin{deluxetable}{cccccccc}
\tabletypesize{\footnotesize}
\tablecaption{Distances to 6.7 GHz methanol masers discovered in AMGPS.\label{table1}}
\tablewidth{0pt}
\tablehead{
\colhead{Source} & \colhead{($v_{min}$, $v_{max}$)} & \colhead{$v_p$} & \colhead{$v_s$} & \colhead{KDA} & \colhead{$d$} & \colhead{$S_i$} & \colhead{$L$} \\
\colhead{} & \colhead{(km s$^{-1}$)} & \colhead{(km s$^{-1}$)} & \colhead{(km s$^{-1}$)} & \colhead{} & \colhead{(kpc)} & \colhead{Jy \kms} & \colhead{($L_\sun$)} 
}
\startdata
34.82+0.35  & 58.5, 60.1 & 59.7 & 57.4 & N & $3.6 \pm 0.4$  &  0.10 & $(9.0 \pm 1.1) \times 10^{-9}$ \\
35.03+0.35  & 40.2, 47.4 & 44.4 & 52.8 & F & $10.4 \pm 0.4$ & 27.41 & $(2.1 \pm 0.2) \times 10^{-5}$ \\
35.25--0.24 & 56.0, 73.3 & 72.4 & 62.0 & N & $3.8 \pm 0.4$  &  0.71 & $(7.1 \pm 1.5) \times 10^{-8}$ \\
35.39+0.02  & 94.0, 97.2 & 96.9 & 95.0 & N & $6.1 \pm 0.7$  &  0.11 & $(2.8 \pm 0.7) \times 10^{-8}$ \\
35.40+0.03  & 88.8, 90.7 & 89.1 & 95.0 & N & $6.1 \pm 0.7$  &  0.30 & $(7.7 \pm 1.8) \times 10^{-8}$ \\
35.59+0.06  & 43.8, 51.8 & 45.9 & 49.1 & F & $10.5 \pm 0.4$ &  0.96 & $(7.3 \pm 0.6) \times 10^{-7}$ \\
35.79--0.17 & 56.6, 64.9 & 60.7 & 60.0 & N & $3.9 \pm 0.4$  & 37.74 & $(4.0 \pm 0.8) \times 10^{-6}$ \\
36.02--0.20 & 92.4, 93.5 & 93.0 & 87.3 & N & $5.4^{+0.8}_{-0.5}$ &  0.08 & $1.6^{+0.5}_{-0.3} \times 10^{-8}$ \\
36.64--0.21 & 77.0, 79.3 & 77.3 & 75.2 & F & $8.9 \pm 0.5$  &  0.52 & $(2.9 \pm 0.4) \times 10^{-7}$ \\
36.70+0.09  & 52.2, 63.2 & 54.7 & 59.7 & F & $9.7 \pm 0.4$  &  8.51 & $(5.5 \pm 0.5) \times 10^{-6}$ \\
36.84--0.02 & 52.8, 64.2 & 61.7 & 59.2 & N & $3.7 \pm 0.4$  &  4.65 & $(4.4 \pm 1.0) \times 10^{-7}$ \\
36.90--0.41 & 83.1, 85.1 & 84.7 & 79.7 & N & $5.0 \pm 0.6$  &  0.26 & $(4.5 \pm 1.1) \times 10^{-8}$ \\
36.92+0.48  & --36.3, --35.6 & --35.9 & --30.7 & F & $15.8 \pm 0.8$ & 0.51 & $(8.8 \pm 0.9) \times 10^{-7}$ \\
37.02--0.03 & 77.5, 85.3 & 78.4 & 80.6 & N & $5.1 \pm 0.6$  &  4.77 & $8.6^{+1.5}_{-1.9} \times 10^{-7}$ \\
37.04--0.04 & 78.1, 86.1 & 84.7 & 80.6 & N & $5.1 \pm 0.6$  &  6.66 & $(1.2 \pm 0.3) \times 10^{-6}$ \\
37.38--0.09 & 67.5, 70.9 & 70.6 & 57.7 & F & $9.7 \pm 0.4$  &  0.13 & $(8.5 \pm 0.8) \times 10^{-8}$ \\
37.47--0.11 & 53.6, 63.3 & 54.7 & 58.8 & F & $9.6 \pm 0.4$  & 22.41 & $(1.4 \pm 0.2) \times 10^{-5}$ \\
37.53--0.11 & 48.2, 56.6 & 50.0 & 53.0 & F & $9.9 \pm 0.4$  &  5.77 & $(3.9 \pm 0.3) \times 10^{-6}$ \\
37.55+0.19  & 78.1, 88.2 & 83.7 & 85.3 & N & $5.6^{+1.1}_{-0.6}$ &  6.44 & $1.4^{+0.6}_{-0.3} \times 10^{-6}$ \\
37.60+0.42  & 84.6, 94.7 & 85.8 & 89.5 & N & $6.2^{+0.5}_{-0.9}$ & 25.29 & $6.7^{+1.2}_{-1.8} \times 10^{-6}$ \\
37.74--0.12 & 49.9, 50.5 & 50.3 & 45.5 & F & $10.3 \pm 0.4$ &  0.25 & $(1.8 \pm 0.2) \times 10^{-7}$ \\
37.76--0.19 & 54.9, 66.0 & 55.1 & 59.2 & F & $9.5 \pm 0.4$  &  1.46 & $(9.1 \pm 0.8) \times 10^{-7}$ \\
37.77--0.22 & 68.8, 70.3 & 69.6 & 61.9 & F & $9.4 \pm 0.5$  &  0.37 & $(2.3 \pm 0.3) \times 10^{-7}$ \\
38.03--0.30 & 54.6, 65.9 & 55.6 & 62.0 & N & $3.9 \pm 0.5$  & 18.10 & $(1.9 \pm 0.5) \times 10^{-6}$ \\
38.08--0.27 & 66.7, 67.8 & 67.5 & 64.6 & N & $4.1 \pm 0.4$  &  0.18 & $(2.1 \pm 0.4) \times 10^{-8}$ \\
38.12--0.24 & 66.5, 79.7 & 70.2 & 83.2 & N & $5.5^{+1.1}_{-0.6}$ &  5.67 & $1.2^{+0.5}_{-0.3} \times 10^{-6}$ \\
38.20--0.08 & 77.7, 88.5 & 79.6 & 83.3 & F & $7.7^{+0.6}_{-1.1}$ & 16.26 & $6.7^{+1.1}_{-1.8}  \times 10^{-6}$ \\
38.26--0.08 &  6.1, 15.9 & 15.4 & 11.6 & F & $12.2 \pm 0.5$ &  5.39 & $(5.6 \pm 0.5) \times 10^{-6}$ \\
38.26--0.20 & 64.1, 73.5 & 70.2 & 65.4 & F & $9.0 \pm 0.5$  &  1.28 & $(7.2 \pm 0.8) \times 10^{-7}$ \\
38.56+0.15  & 23.1, 31.2 & 31.5 & 29.3 & N & $2.1 \pm 0.5$  &  0.13 & $4.0^{+2.0}_{-1.7} \times 10^{-9}$ \\
38.60--0.21 & 61.4, 69.5 & 62.6 & 66.4 & N & $4.2 \pm 0.5$  &  0.59 & $(7.2 \pm 1.8) \times 10^{-8}$ \\
38.66+0.08  & --31.9, --30.7 & --31.5 & --39.2 & F & $16.3 \pm 0.9$ & 0.71 & $(1.3 \pm 0.2) \times 10^{-6}$ \\
38.92--0.36 & 30.8, 33.5 & 31.9 & 38.7 & F & $10.5 \pm 0.4$ &  1.29 & $(9.9 \pm 0.8) \times 10^{-7}$ \\
39.39--0.14 & 58.2, 75.5 & 60.4 & 66.1 & N & $4.3 \pm 0.5$  &  0.77 & $(9.9 \pm 2.3) \times 10^{-8}$ \\
39.54--0.38 & 47.4, 49.4 & 47.8 & 60.9 & F & $9.0 \pm 0.5$  &  0.24 & $(1.3 \pm 0.2) \times 10^{-7}$ \\
40.28--0.22 & 62.4, 85.7 & 73.9 & 73.0 & N & $4.9^{+0.9}_{-0.6}$ & 68.03 & $1.1^{+0.5}_{-0.2} \times 10^{-5}$ \\
40.62--0.14 & 29.7, 36.7 & 31.1 & 32.3 & F & $10.5 \pm 0.4$ &  6.97 & $(5.3 \pm 0.4) \times 10^{-6}$ \\
40.94--0.04 & 36.2, 43.2 & 36.6 & 40.2 & F & $10.0 \pm 0.5$ &  1.50 & $(1.0 \pm 0.1) \times 10^{-6}$ \\
41.08--0.13 & 57.2, 58.4 & 57.5 & 63.8 & F & $8.4 \pm 0.6$  &  0.33 & $(1.6 \pm 0.3) \times 10^{-7}$ \\
41.12--0.11 & 33.1, 37.4 & 36.6 & 38.1 & F & $10.0 \pm 0.5$ &  0.63 & $(4.4 \pm 0.5) \times 10^{-7}$ \\
41.12--0.22 & 55.0, 66.6 & 63.4 & 60.1 & F & $8.7 \pm 0.6$  &  1.23 & $(6.5 \pm 0.9) \times 10^{-7}$ \\
41.16--0.20 & 61.6, 63.8 & 63.6 & 59.9 & F & $8.7 \pm 0.6$  &  0.20 & $(1.0 \pm 0.2) \times 10^{-7}$ \\
41.23--0.20 & 54.0, 64.9 & 55.4 & 59.2 & F & $8.7 \pm 0.5$  &  7.55 & $(4.0 \pm 0.5) \times 10^{-6}$ \\
41.27+0.37  & 19.4, 20.6 & 20.3 & 14.6 & F & $11.5 \pm 0.5$ &  0.16 & $(1.5 \pm 0.2) \times 10^{-7}$ \\
41.34--0.14 &  6.6, 15.0 & 11.7 & 12.6 & F & $11.6 \pm 0.5$ & 25.40 & $(2.4 \pm 0.2) \times 10^{-5}$ \\
41.58+0.04  & 10.4, 12.3 & 11.9 &  12.4& F & $11.5 \pm 0.5$ &  0.22 & $(2.0 \pm 0.2) \times 10^{-7}$ \\
41.87--0.10 & 15.5, 23.7 & 15.8 & 18.8 & F & $11.1 \pm 0.5$ &  0.10 & $(8.5 \pm 0.8) \times 10^{-8}$ \\
42.03+0.19  &  6.8, 17.3 & 12.8 & 18.0 & F & $11.1 \pm 0.5$ & 30.10 & $(2.6 \pm 0.3) \times 10^{-5}$ \\
42.30--0.30 & 26.2, 34.7 & 28.1 & 27.1 & F & $10.5 \pm 0.5$ &  5.07 & $(3.9 \pm 0.4) \times 10^{-6}$ \\
42.43--0.26 & 65.8, 69.1 & 66.8 & 64.5 & F & $7.9 \pm 0.8$  &  1.61 & $(7.0 \pm 1.4) \times 10^{-7}$ \\
42.70--0.15 & --47.1, --39.0 & --42.9 & --44.3 & F & $15.9 \pm 0.8$ & 4.03 & $(7.1 \pm 0.7) \times 10^{-6}$ \\
43.04--0.46 & 54.1, 63.6 & 54.8 & 57.3 & F & $8.3 \pm 0.6$  & 10.39 & $(5.0 \pm 0.7) \times 10^{-6}$ \\
43.08--0.08 &  9.6, 14.9 & 10.2 & 12.3 & F & $11.2 \pm 0.5$ &  4.11 & $(3.6 \pm 0.3) \times 10^{-6}$ \\
43.15+0.02  & 12.3, 14.3 & 13.3 & 10.9 & F & $11.4 \pm 1.2$ & 12.03 & $(1.1 \pm 0.2) \times 10^{-5}$ \\
43.16+0.02  &  6.8, 22.1 &  9.3 &  3.5 & F & $11.4 \pm 1.2$ & 50.54 & $(4.6 \pm 1.0) \times 10^{-5}$ \\
43.17+0.01  & 18.1, 22.4 & 19.0 & 12.0 & F & $11.4 \pm 1.2$ & 15.05 & $(1.4 \pm 0.3) \times 10^{-5}$ \\
43.17--0.00 & --1.7, 4.2 & --1.2 & 1.7 & F & $11.4 \pm 1.2$ &  1.44 & $(1.3 \pm 0.3) \times 10^{-6}$ \\
43.18--0.01 & 10.3, 11.6 & 11.1 & 11.9 & F & $11.4 \pm 1.2$ &  0.63 & $(5.7 \pm 1.2) \times 10^{-7}$ \\
43.80--0.13 & 38.4, 43.6 & 39.6 & 43.7 & F & $9.1 \pm 0.5$  & 43.18 & $(2.5 \pm 0.3) \times 10^{-5}$ \\
44.31+0.04  & 55.0, 56.6 & 55.7 & 57.0 & F & $7.9 \pm 0.7$  &  0.43 & $(1.9 \pm 0.4) \times 10^{-7}$ \\
44.64--0.52 & 48.8, 49.9 & 49.3 & 46.0 & F & $8.7 \pm 0.5$  &  0.29 & $(1.5 \pm 0.2) \times 10^{-7}$ \\
45.07+0.13  & 56.8, 60.0 & 57.8 & 59.2 & F & $7.4 \pm 0.8$  & 22.26 & $(8.4 \pm 1.6) \times 10^{-6}$ \\
45.44+0.07  & 49.1, 50.6 & 50.0 & 58.5 & F & $7.4^{+0.6}_{-1.5}$ & 0.71 & $2.7^{+0.4}_{-1.0} \times 10^{-7}$ \\
45.47+0.05  & 55.4, 59.8 & 56.0 & 59.7 & F & $7.2^{+0.7}_{-1.3}$ & 9.40 & $3.4^{+0.7}_{-1.1} \times 10^{-6}$ \\
45.47+0.13  & 57.0, 73.5 & 65.7 & 61.9 & F & $6.9^{+0.8}_{-1.0}$ & 4.88 & $(1.6 \pm 0.4) \times 10^{-6}$ \\
45.49+0.13  & 56.7, 66.4 & 57.2 & 60.5 & F & $7.1^{+0.7}_{-1.2}$ & 4.98 & $(1.7 \pm 0.5) \times 10^{-6}$ \\
45.57--0.12 &  1.2, 9.8  &  1.6 &  4.8 & F & $11.2 \pm 0.5$ &  0.30 & $(2.6 \pm 0.2) \times 10^{-7}$ \\
45.81--0.36 & 54.7, 70.8 & 59.9 & 60.3 & F & $7.0^{+0.8}_{-1.1}$ & 8.41 & $2.9^{+0.6}_{-0.9} \times 10^{-6}$ \\
46.07+0.22  & 22.3, 25.1 & 23.3 & 19.5 & F & $10.1 \pm 0.5$ &  1.20 & $(8.5 \pm 0.8) \times 10^{-7}$ \\
46.12+0.38  & 57.5, 62.9 & 59.0 & 55.0 & F & $7.5^{+0.6}_{-1.0}$ & 1.05 & $4.1^{+0.7}_{-1.0} \times 10^{-7}$ \\
48.89--0.17 & 57.2, 57.5 & 57.3 & 55.7 & F & $5.1^{+2.9}_{-1.4}$  &  0.02 & $3.6^{+5.3}_{-1.6} \times 10^{-9}$ \\
48.90--0.27 & 63.6, 72.5 & 72.0 & 68.4 & F & $5.1^{+2.9}_{-1.4}$  &  0.57 & $1.0^{+1.5}_{-0.5} \times 10^{-7}$ \\
48.99--0.30 & 62.5, 72.6 & 71.6 & 67.5 & F & $5.1^{+2.9}_{-1.4}$  &  0.55 & $1.0^{+1.4}_{-0.5} \times 10^{-7}$ \\
49.27+0.31  & --6.8, 7.5 & --3.2 & 3.4 & F & $10.5 \pm 0.5$ & 13.71 & $(1.0 \pm 0.1) \times 10^{-5}$ \\
49.35+0.41  & 66.1, 69.2 & 68.0 & 65.5 & F & $5.1^{+2.9}_{-1.4}$  &  6.72 & $1.2^{+1.8}_{-0.6} \times 10^{-6}$ \\
49.41+0.33  & --27.0, --9.8 & --12.1 & --21.3 & F & $12.2 \pm 0.6$ & 23.18 & $(2.4 \pm 0.2) \times 10^{-5}$ \\
49.47--0.37 & 55.6, 76.1 & 63.8 & 62.4 & F & $5.1^{+2.9}_{-1.4}$  & 17.35 & $3.1^{+4.6}_{-1.5} \times 10^{-6}$ \\
49.48--0.40 & 47.7, 65.3 & 51.4 & 58.2 & F & $5.1^{+2.9}_{-1.4}$  & 14.90 & $2.7^{+3.9}_{-1.3} \times 10^{-6}$ \\
49.49--0.37 & 54.4, 65.7 & 56.1 & 60.6 & F & $5.1^{+2.9}_{-1.4}$  & 49.27 & $0.9^{+1.3}_{-0.4} \times 10^{-5}$ \\
49.49--0.39 & 49.9, 65.0 & 59.3 & 56.6 & F & $5.1^{+2.9}_{-1.4}$  & 574.7 & $1.0^{+1.5}_{-0.5} \times 10^{-4}$ \\
49.60--0.25 & 59.9, 66.7 & 62.9 & 57.2 & F & $5.1^{+2.9}_{-1.4}$  & 78.16 & $1.4^{+2.1}_{-0.7} \times 10^{-5}$ \\
49.62--0.36 & 48.8, 60.1 & 49.3 & 54.9 & F & $5.1^{+2.9}_{-1.4}$  &  1.13 & $2.0^{+3.0}_{-0.9} \times 10^{-7}$ \\
50.78+0.15  & 47.6, 50.9 & 49.1 & 42.2 & F & $7.0^{+0.7}_{-1.1}$ & 2.79 & $9.5^{+3.0}_{-2.8} \times 10^{-7}$ \\
52.92+0.41  & 38.8, 45.0 & 39.1 & 44.8 & T & 5.1            &  4.68 & $8.4 \times 10^{-7}$ \\
53.04+0.11  &  9.7, 10.5 & 10.1 &  5.5 & F & $9.4 \pm 0.5$  &  0.69 & $(4.2 \pm 0.5) \times 10^{-7}$ \\
53.14+0.07  & 23.4, 25.4 & 24.6 & 21.8 & N & $1.9 \pm 0.6$  &  0.54 & $(1.4 \pm 0.9) \times 10^{-8}$ \\
53.62+0.04  & 18.2, 19.5 & 19.0 & 23.3 & N & $2.1 \pm 0.6$  &  7.16 & $2.2^{+1.4}_{-1.1} \times 10^{-7}$ \\
\enddata
\tablecomments{The columns show the source name, range of maser emission ($v_{min}$, $v_{max}$), velocity of peak maser emission, $v_p$, systemic velocity from molecular line measurements, $v_s$, resolution of the KDA - near (N), far (F) or tangent point (T), distance, $d$, integrated flux, $S_i$, and isotropic luminosity, $L$ (calculated from $S_i$). Asymmetric uncertainties in the luminosity are typically a consequence of the asymmetric uncertainties in the distance to the source.}
\end{deluxetable}

\begin{table}
\begin{center}
\caption{Number of 6.7 GHz methanol masers that should have been detected in extragalactic surveys to date if the maser populations in the galaxies were similar to that in the Milky Way.\label{table2}}
\begin{tabular}{cccc}
\tableline\tableline
Galaxy & Limiting $S_p$ & \multicolumn{2}{c}{Number of detectable masers} \\
 & (mJy) & Method-I & Method-II \\
\tableline
LMC & 270 & 180 & 195 \\
SMC & 400 & 100 & 110 \\
M33 &  4  &  70 &  90 \\
\tableline
\end{tabular}
\end{center}
\end{table}

\clearpage
{\epsscale{0.9}\plotone{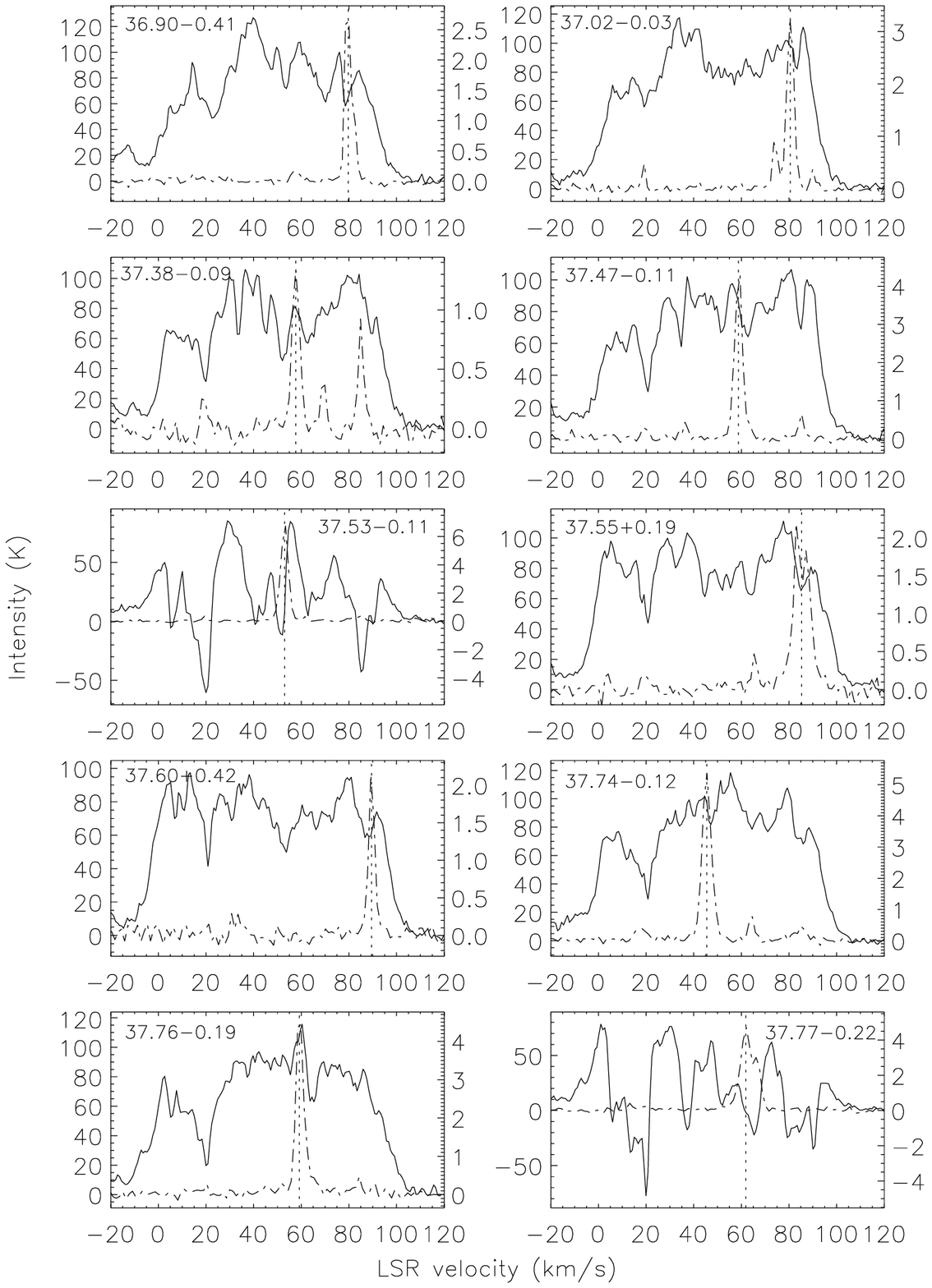}}\\[5mm]
\centerline{Fig.~1. --- Continued.}
\clearpage
{\epsscale{0.9}\plotone{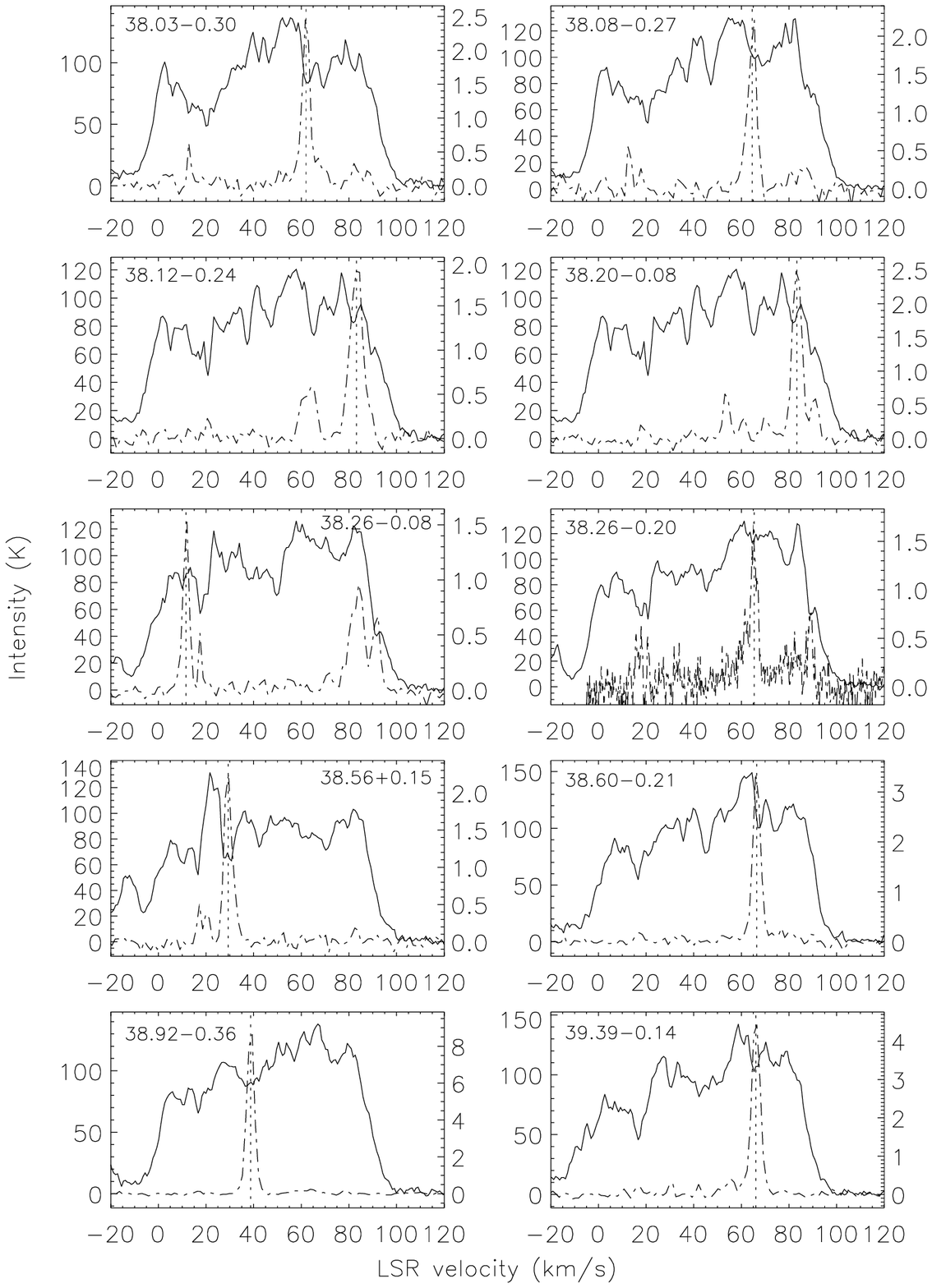}}\\[5mm]
\centerline{Fig.~1. --- Continued.}
\clearpage
{\epsscale{0.9}\plotone{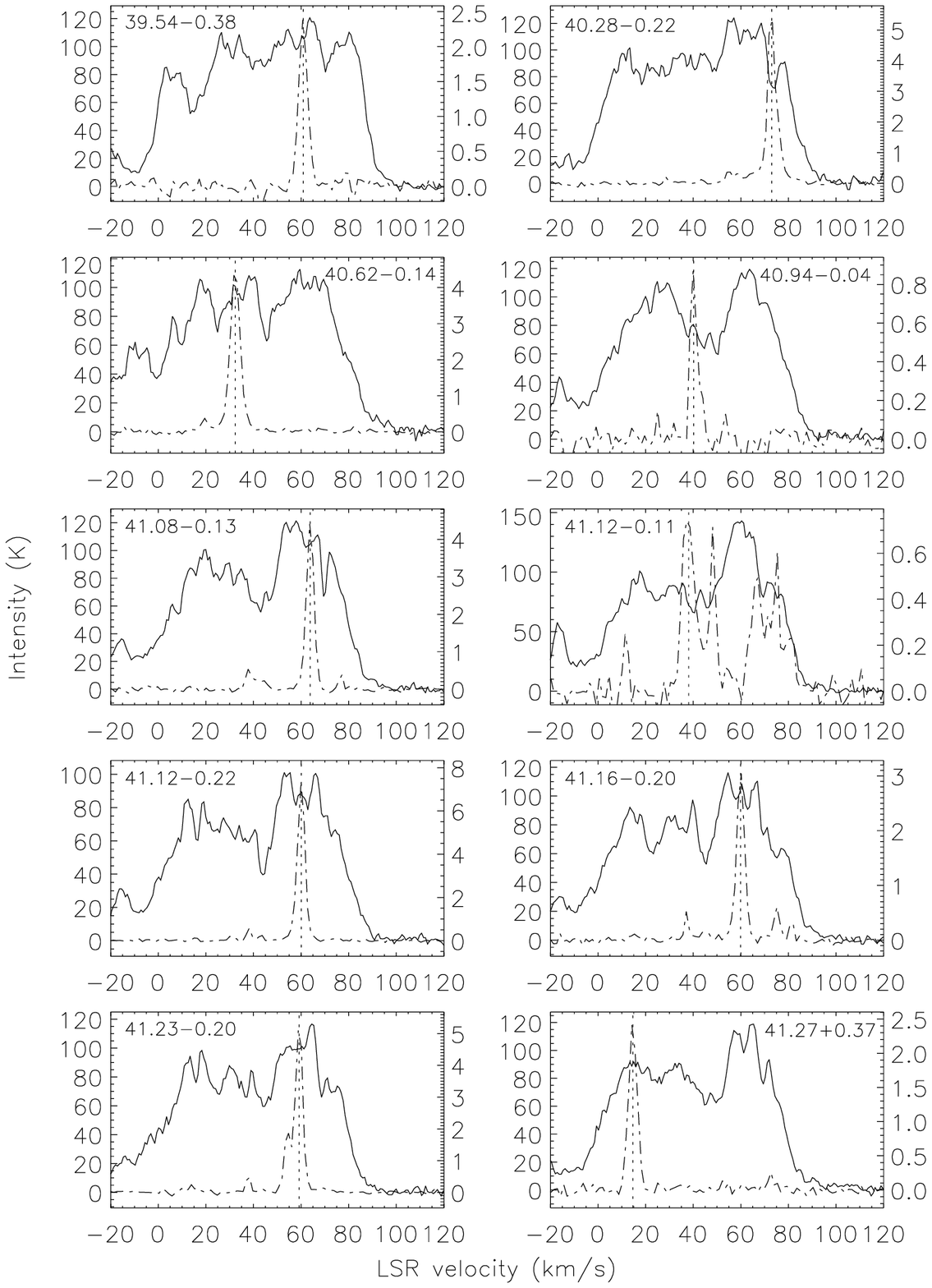}}\\[5mm]
\centerline{Fig.~1. --- Continued.}
\clearpage
{\epsscale{0.9}\plotone{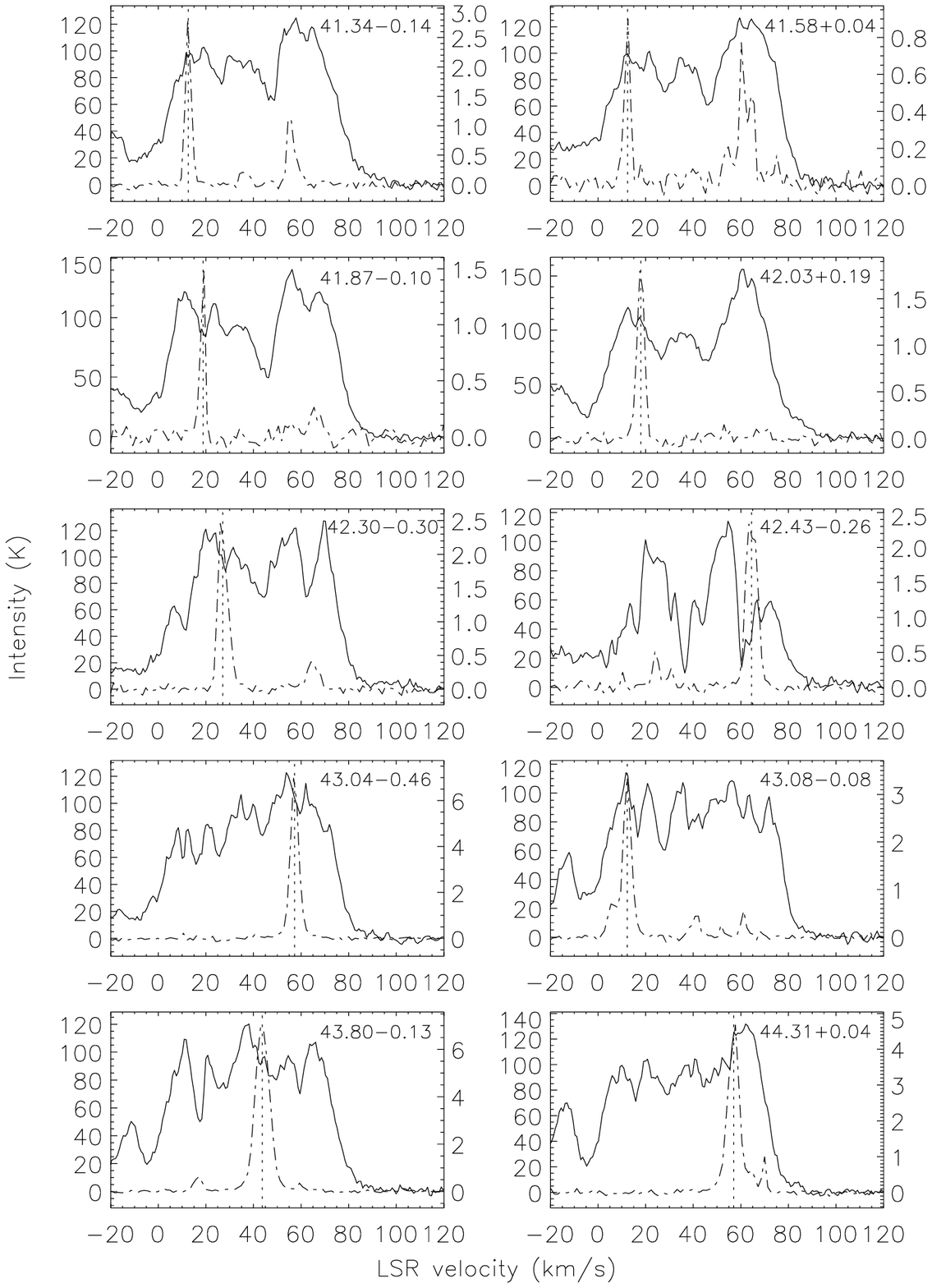}}\\[5mm]
\centerline{Fig.~1. --- Continued.}
\clearpage
{\epsscale{0.9}\plotone{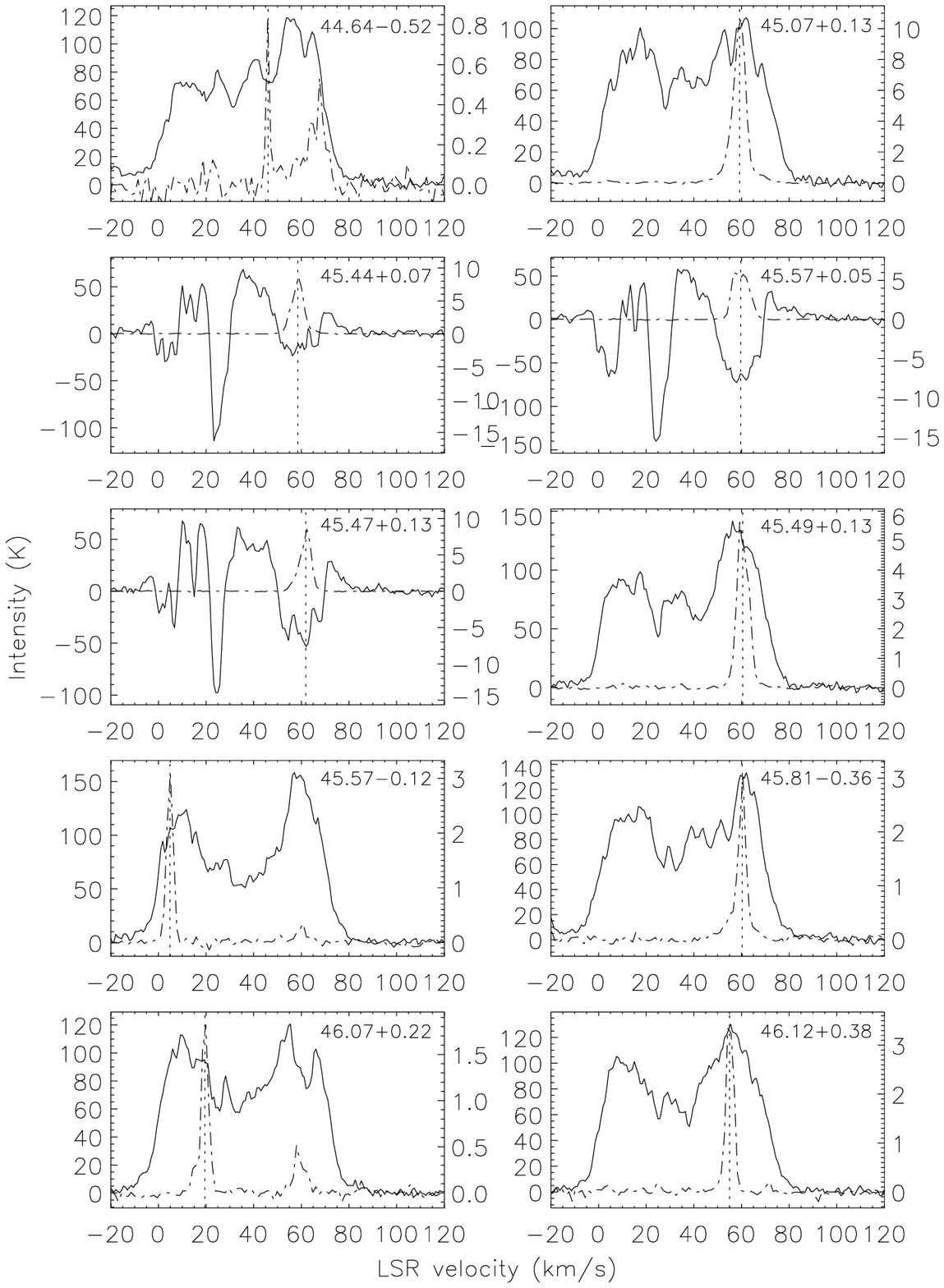}}\\[5mm]
\centerline{Fig.~1. --- Continued.}
\clearpage
{\epsscale{0.9}\plotone{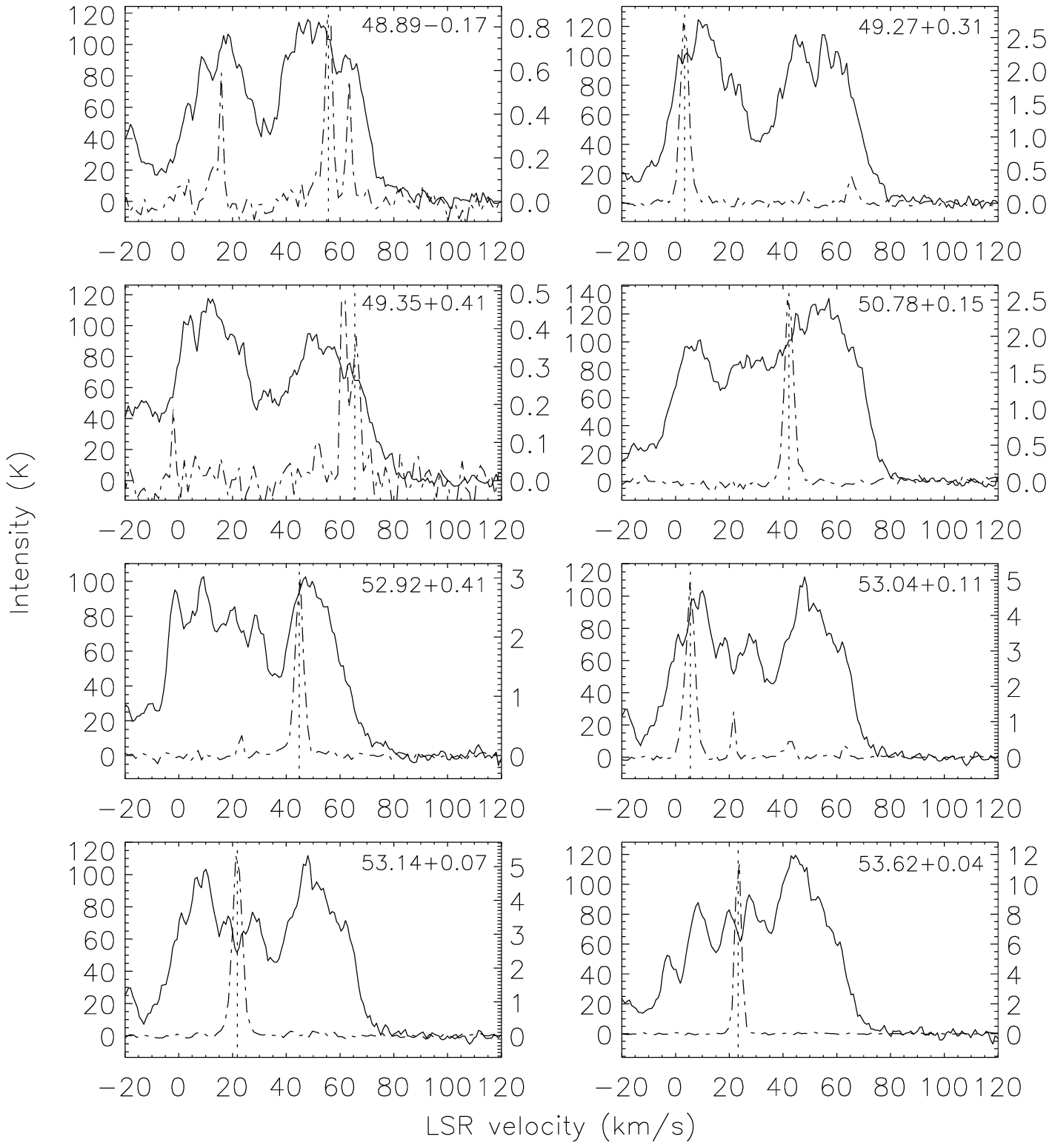}}\\[5mm]
\centerline{Fig.~1. --- Continued.}
\clearpage

\end{document}